\documentclass{aa}  
\usepackage{graphicx}
\usepackage[varg]{txfonts}
\usepackage{natbib}
\bibpunct{(}{)}{;}{a}{}{,} 
\newcommand{\formaldehyde}{$\mathrm{H_2CO}$}

\newcommand{\methanol}{$\mathrm{CH_3OH}$}
\newcommand{\htwo}{$\mathrm{H_2}$}
\newcommand{\gstate}{$1_{10}-1_{11}$}
\newcommand{\fstate}{$2_{11}-2_{12}$}
\newcommand{\sstate}{$3_{12}-3_{13}$}
\newcommand{\tstate}{$4_{13}-4_{14}$}
\newcommand{\taug}{$\tau_{4.8}$}
\newcommand{\taumax}{$\tau_{4.8}$(max)}
\newcommand{\nhtwo}{$n_{H_2}$}
\newcommand{\tk}{$T_k$}
\newcommand{\td}{$T_d$}
\newcommand{\wdust}{$W_d$}
\newcommand{\whii}{$W_{\ion{H}{II}}$}
\begin{document} 
\title{Pumping of the 4.8 GHz H$_{\text{2}}$CO masers and its implications for
  the periodic masers in G37.55+0.20} \subtitle{} \titlerunning{Periodic masers
  in G37.55+0.20E} \author{D.J. van der Walt \inst{1}} \institute{Center for
  Space Research, Potchefstroom Campus, North-West University, Potchefstroom,
  South Africa\\ \email{johan.vanderwalt@nwu.ac.za}}

\date{}

 \abstract{Periodic or regular flaring of class II \methanol{} masers in nine high mass
   star forming regions is now a well established phenomenon. Amongst the nine star
   forming regions, G37.55+0.20 is the only case at present where apart from the presence
   of a periodic class II \methanol{} maser, correlated flaring of another masing species,
   \formaldehyde{} in this case, has been detected.}  {We perform numerical calculations
   to investigate under which conditions the \gstate{} transition (4.8 GHz) of
   ortho-\formaldehyde{} is inverted in order to address the question of the correlated
   flaring of the 6.7 GHz \methanol{} and 4.8 GHz \formaldehyde{} masers in G37.55+0.20}
          {We developed a numerical code to study the population inversion of
            o-\formaldehyde{}. Equilibrium solutions for the level populations are found
            by integrating the rate equations using Heun's method.}  {It is found that
            collisional excitation with \htwo{} as well as radiative excitation by the
            free-free radio continuum radiation from a nearby ultra- or hyper-compact
            \ion{H}{II} region can invert the \gstate{} transition. It is also found that
            radiative excitation by the dust infrared radiation field does not lead to an
            inversion of the \gstate{} transition. The \fstate{} (14.5 GHz) and \sstate{}
            (28.9 GHz) transitions are inverted only in the presence of the free-free
            continuum radiation field of a very compact \ion{H}{II} region.} {Due to the
            different pumping mechanisms of the \formaldehyde{} and \methanol{} masers it
            is unlikely that the near simultaneous flaring of the \methanol{} and
            \formaldehyde{} masers in G37.55+0.20 is due to changes in the pumping of the
            masers.}   \maketitle
%

\section{Introduction}
The phenomenon of periodic or regular flaring of class II \methanol{} masers associated
with a small number of high mass star forming regions is now well established and is
considered as one of the most peculiar phenomena in maser science
\citep{menten2012}. Periodic or regular flaring behaviour of 6.7, 12.2, and 107 GHz
\methanol{} masers have been reported and discussed by \citet{goedhart2003, goedhart2007,
  goedhart2009}, \citet{vanderwalt2009}, \citet{araya2010} and
\citet{szymczak2011}. Currently nine confirmed periodic maser sources are known.

Since the original discovery of periodic flaring of the 6.7 GHz \methanol{} masers in
G9.62+0.20E by \citet{goedhart2003}, numerous suggestions/hypotheses have been proposed
about the underlying mechanisms responsible for the periodic behaviour of the masers. The
underlying mechanisms can be divided into two groups, ie. radiative mechanisms that affect
the pumping of the masers and, background source effects which affect the flux of seed
photons amplified by the maser.

Amongst the nine known periodic \methanol{} maser sources, G37.55+0.20 (IRAS 18566+0408)
is at present the only where another masing species, \formaldehyde{} in this case, shows
flaring behaviour which occurs almost coincident in time with the \methanol{} maser flares
\citep{araya2010}. Based on the fact that both species flare and that \methanol{} is
radiatively pumped \citep{sobolev1994,sobolev1997}, \citet{araya2010} concluded that the
\formaldehyde{} maser in G37.55+0.20 must also be radiatively pumped. These authors
therefore proposed an underlying mechanism which raises the dust temperature and which in
turn affects the infrared pumping radiation field, resulting in an increase in the
amplification of the masers.

Level population calculations for \formaldehyde{} have been done in the past by
eg. \citet{thaddeus1972}, \citet{boland1981}, and \citet{baan1986}. \citet{boland1981}
presented a model in which the \formaldehyde{} masers in NGC 7538-IRS1 are radiatively
pumped by the free-free radio continuum radiation of a nearby compact \ion{H}{II}
region. A number of authors eg. \citet{mehringer1994} and \citet{hoffman2003}, however,
argued that the Galactic 4.8 GHz \formaldehyde{} maser emission is due to collisional
rather than radiative excitation from dust emission. On the other hand, numerical
calculations strongly suggest that the class II \methanol{} masers are pumped by infrared
radiation \citep{sobolev1994,sobolev1997} which obviously raises the question of the
underlying mechanism that apparently drives both the \methanol{} and \formaldehyde{} maser
flares in G37.55+0.20.

Thus, although the pumping of the Galactic 4.8 GHz \formaldehyde{} masers is an
interesting problem as such, the near simultaneous flaring of the 6.7 GHz \methanol{} and
4.8 GHz \formaldehyde{} masers in G37.55+0.20 puts the question of the pumping of the
\formaldehyde{} masers in a new context. In fact, careful comparison of the 6.7 GHz
\methanol{} and 4.8 GHz \formaldehyde{} flare profiles as presented by \citet{araya2010}
reveals a remarkable similarity between the two. This also again raises the question of
the pumping of the 4.8 GHz \formaldehyde{} masers and what the implication thereof is for
identifying the underlying mechanism responsible for the \methanol{} and \formaldehyde{}
maser flares in G37.55+0.20 especially when considering that the two types of masers are
2000 AU apart \citep{araya2010}. The aim of this paper is to present the results of an
investigation into the pumping of the 4.8 GHz \formaldehyde{} masers and to use the
results as a step to understand the underlying mechanism for the flaring of the masers in
G37.55+0.20 and hopefully also of the other periodic maser sources.

\section{Theory and Calculational procedure}
 
Within the framework of the escape probability method the rate equations for the
level populations can be written as 
\begin{eqnarray}\nonumber
\frac{dN_i}{dt} & = & \sum_{j < i}[(-N_i + (\frac{g_i}{g_j}N_j -
  N_i)W\mathcal{N}_{ij})\beta_{ij}A_{ij} \\  
\nonumber & &  + C_{ij}(N_j\frac{g_i}{g_j}e^{-E_{ij}/{kT}} - N_i)]\\
\nonumber  & & +  \sum_{j > i}(N_j + (N_j  -
\frac{g_j}{g_i}N_i)W\mathcal{N}_{ji})\beta_{ji}A_{ji} \\ 
& & + C_{ji}(N_j - N_i\frac{g_j}{g_i}e^{-E_{ji}/{kT}})]
\label{eq:rate}
\end{eqnarray}
where $N_i$ is the number density in level $i$, $g_i$ the statistical weight of
level $i$, $A_{ij}$ the Einstein-A coefficient for spontaneous emission between
levels $i$ and $j$, $W$ the dilution factor for an external radiation field and
$\mathcal{N}_{ij}$ the photon occupation number for this field at frequency
$\nu_{ij}$. $C_{ij} = n_{H_2} K_{ij}$ is the collision rate with $n_{H_2}$ the
\htwo{} number density and $K_{ij}$ the collision rate coefficient. $\beta_{ij}$
is the escape probability. The rate equations were supplemented with the
particle number conservation requirement
\begin{equation}
N_{tot} = Xn_{H_2} = \sum_i N_i
\end{equation}
with $X$ the abundance of the molecule under consideration relative to that of
\htwo{}.

We assume that the maser region is embedded in or surrounded by a layer of warm dust at
temperature \td. For the dust radiation field we assume the isotropic optically thin case
and therefore that the dust radiation field is given by
\begin{equation}
I_{\nu} = W_d\tau_d(\nu)B_\nu(T_d)
\end{equation}
with $B_\nu(T_d)$ the Planck function and $\tau_d(\nu) =
\tau_d(\nu_0)(\nu/\nu_0)^{-p}$ the dust optical depth where we have set
$\tau_d(\nu_0=10^{13}\,\mathrm{Hz}) = 1$ and $p = 2$ \citep[see
  eg.][]{pavlakis1996,sobolev1997,cragg2002}. $W_d \leq 1$ is a geometric dilution factor.

The photon occupation number, $\mathcal{N}_{ij}$, enters in the derivation of
Eq. \ref{eq:rate} through the product of $c^2/h\nu^3$ with the mean intensity of
the continuum radiation field averaged over the line profile. For dust emission
it is given
by
\begin{equation}
\mathcal{N}_{ij,d} = W_d(\frac{\nu_{ij}}{\nu_0})^{-p}\frac{1}{e^{h\nu_{ij}/kT_d} -1}
\label{eq:dust}
\end{equation}
while for the radiation field of a nearby \ion{H}{II} region it is given by 
\begin{equation}
\mathcal{N}_{ij,\ion{H}{II}} = W_{\ion{H}{II}}\frac{1 - 
  e^{-\tau_{\nu_{ij}}}}{e^{h\nu_{ij}/kT_e} -1}
\end{equation}
where $T_e$ is the electron temperature, $\tau_{\nu_{ij}}$ the optical depth of the
\ion{H}{II} region at frequency $\nu_{ij}$, and $W_{\ion{H}{II}}$ the corresponding
geo\-me\-tric dilution factor. In the case where the continuum radiation field consists of
both that of dust and an \ion{H}{II} region the product $W\mathcal{N}_{ij}$ in
Eq. \ref{eq:rate} is to be replaced by $W_d\mathcal{N}_{ij,d} +
W_{\ion{H}{II}}\mathcal{N}_{ij,\ion{H}{II}}$.

The optical depth between levels $i$ and $j$ is given by
\begin{equation}
\tau_{ij} =
\frac{A_{ij}}{8\pi}\left(\frac{c}{\nu}\right)^3\left(\frac{g_j}{g_i}x_i -
x_j\right)\frac{N_{col}}{\Delta v}
\label{eq:tau}
\end{equation}
where $x_i$ and $x_j$ are the fractional number of molecules in levels $i$ and $j$
respectively, $N_{col} = Xn_{H_2}d$ the column density of the masing molecule under
consideration with $d$ the source dimension that determines the photon escape, and $\Delta
v$ the FWHM of the local velocity field. Following \citet{cragg2002,cragg2005} we will
refer to the quantity $N_{col}/\Delta v$ as the specific column density.  For the escape
probability we used the expression for the large velocity gradient approximation, ie.
\begin{equation}
\beta_{\mathrm{LVG}} = \frac{1 - e^\tau}{\tau}.
\end{equation} 

The set of rate equations was solved using Heun's method \citep{kreyszig} with a time step
of one second. Using Heun's method requires an initial distribution for the level
populations which we took as a Boltzmann distribution with a temperature equal to the gas
kinetic temperature (henceforth indicated by \tk). For a given \nhtwo{} the calculation
always started with a small \formaldehyde{} specific column density such that the system
is completely optically thin in all radiative transitions.  An equilibrium solution for
this initial specific column density was then found by letting the system evolve over a
long enough time. The specific column density is then increased by a small amount and the
equilibrium solution for the previous value of the specific column density is then used as
the initial distribution. The process is repeated until the required specific column
density is reached. Experimentation with this procedure has shown that starting with the
optically thin case and progressively calculating solutions for increasing optical depth
is very stable and repeatable.

In applying the above method in more extensive calculations it is not practical to find
the equilibrium level populations by letting the system evolve over, eg. $\geq 10^6$
seconds. Instead we used a convergence criterium to determine when to stop the
calculation. For the present calculations it was required that the maximum value of the
relative change in one time step over {\it all} levels be less than $10^{-7}$,
ie. $|N_i(t_{j+1})-N_i(t_j)|/N_i(t_j) < 10^{-7}$ for all $i$. From a number of trial runs
it was found that when using this convergence criterium, the level populations differ by
less than 1\% from the equilibrium solutions found when allowing the system to evolve over
a sufficiently long time ($>10^6$ seconds).

The Einstein-A and collision rate coefficients of the first 40 levels of
o-\formaldehyde{} were taken from the Leiden Atomic and Molecular Database
\citep{lamda2005}. The \htwo-\formaldehyde{} collision rate coefficients used
are from the recent calculations of \citet{wiesenfeld2013}.

The calculational procedure was tested by considering the two limiting cases where (a)
collisions dominate and, (b) where the radiation field dominates and is an undiluted black
body radiation field. In the first case it is expected that in equilibrium the level
populations be described by a Boltzmann distribution with temperature equal to the gas
kinetic temperature and in the second case by a Boltzmann distribution with temperature
equal to that of the black body. In both these cases it was found that the equilibrium
level populations calculated with the above method produced the correct temperatures with
an accuracy of less than one per cent. 

\section{Results}

There are basically three processes in the physical environment that can play a role in
the pumping of the masers, ie. collisional excitation, radiative excitation via the dust
continuum emission, and radiative excitation through the free-free continuum emission from
an ultra- or hypercompact \ion{H}{II} region. We consider each of these separately. The
focus will be on the \gstate{} transition but we will also later present results for other
possible masing transitions.

\subsection{Collisional excitation}

In the absence of radiation fields the effect of collisions on the level populations is
determined by \tk{}, \nhtwo{}, and the relative abundance of \formaldehyde{},
$X_{H_2CO}$. We do not consider explicitly the dependence on the relative
abundance of \formaldehyde{} but implicitly through the \formaldehyde{} specific column
density, $X_{H_2CO}n_{H_2}\ell/\Delta v$ (Eq. \ref{eq:tau}). For illustrative
purposes we considered gas kinetic temperatures in steps of 40 K between 140 K and 300 K
and \htwo{} densities between $10^4~\mathrm{cm^{-3}}$ and $3\times 10^7~\mathrm{cm^{-3}}$.

\begin{figure}[h]
   \centering \includegraphics[width=9cm]{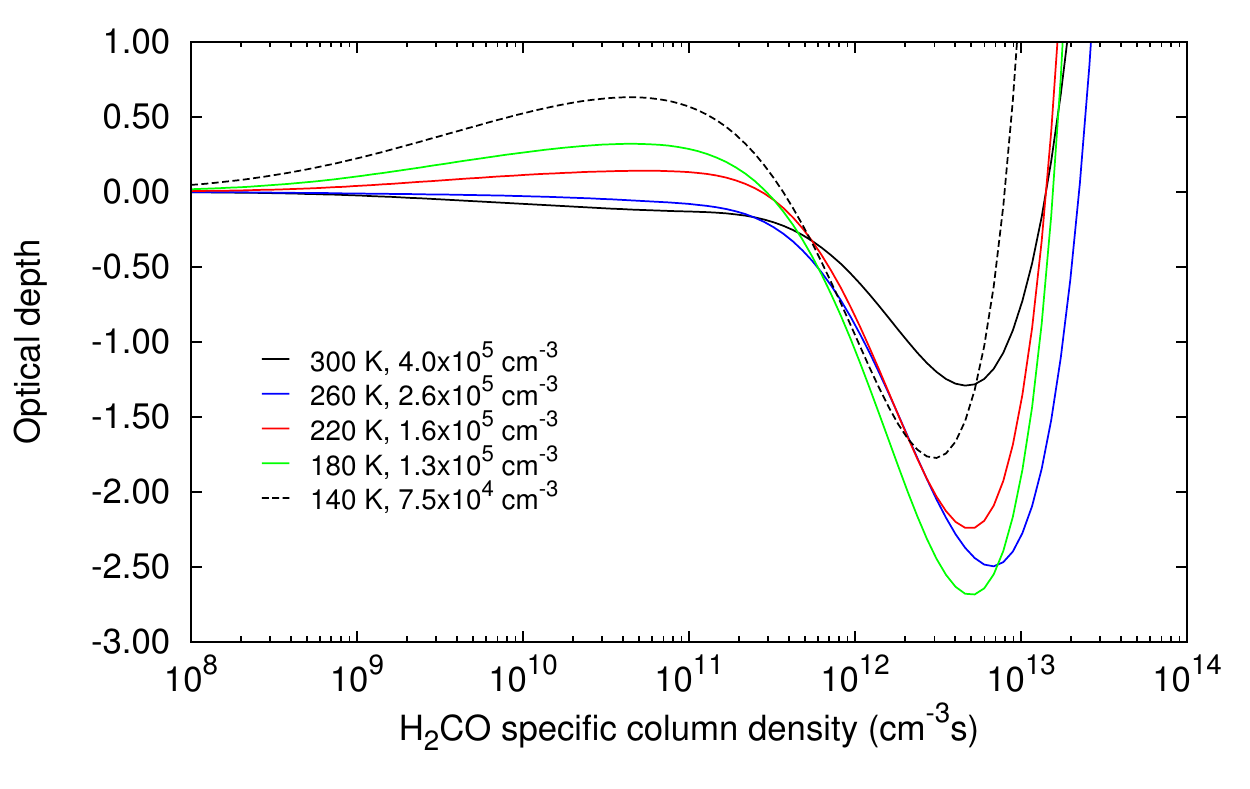}
   \caption{Comparison of the dependence of \taug{} on the \formaldehyde{} specific column
     density for five combinations of \tk{} and \nhtwo{}. No radiative excitation was considered.}
   \label{fig:tauvsncol1}
  \end{figure}

In Fig. \ref{fig:tauvsncol1} we show the optical depth, \taug{}, of the \gstate{}
transition as a function of the \formaldehyde{} specific column density for five values of
\tk{}. The calculation was done for a specific value of \nhtwo{} for each value of
\tk{}. The choices for the densities will be explained below. It is seen that for the
combinations of lower temperatures and \htwo{} densities there is no inversion below
specific column densities of about $3 \times 10^{11}~\mathrm{cm^{-3}\,s}$. For the higher
temperatures, ie. 260 K and 300 K, an inversion occurs already from the lowest considered
specific column densities with the resulting optical depth being very small (\taug{} $\ll
1$) for specific column densities less than about $3 \times
10^{11}~\mathrm{cm^{-3}\,s}$. Significant amplification for all five combinations of \tk{}
and \nhtwo{} occurs only for specific column densities greater than about
$10^{12}~\mathrm{cm^{-3}\,s}$.

To investigate and illustrate the behaviour of \taug{} for more general combinations of
\tk{} and \nhtwo{} in terms of the dependence of \taug{} on the specific column density
would require a large number of graphs such as in Fig. \ref{fig:tauvsncol1} which is
unpractical. Instead we will characterize the dependence of \taug{} on the specific column
density for a single combination of \tk{} and \nhtwo{} with \taumax{}, the maximum
negative value of \taug{} over the considered range of specific column densities.  The
detailed information of the dependence of \taug{} on specific column density for a given
combination of \tk{} and \nhtwo{} is therefore lost. However, \taumax{} is still a usefull
quantity to investigate under which conditions a population inversion occurs.

\begin{figure}[h]
     \centering
     \includegraphics[width=9cm]{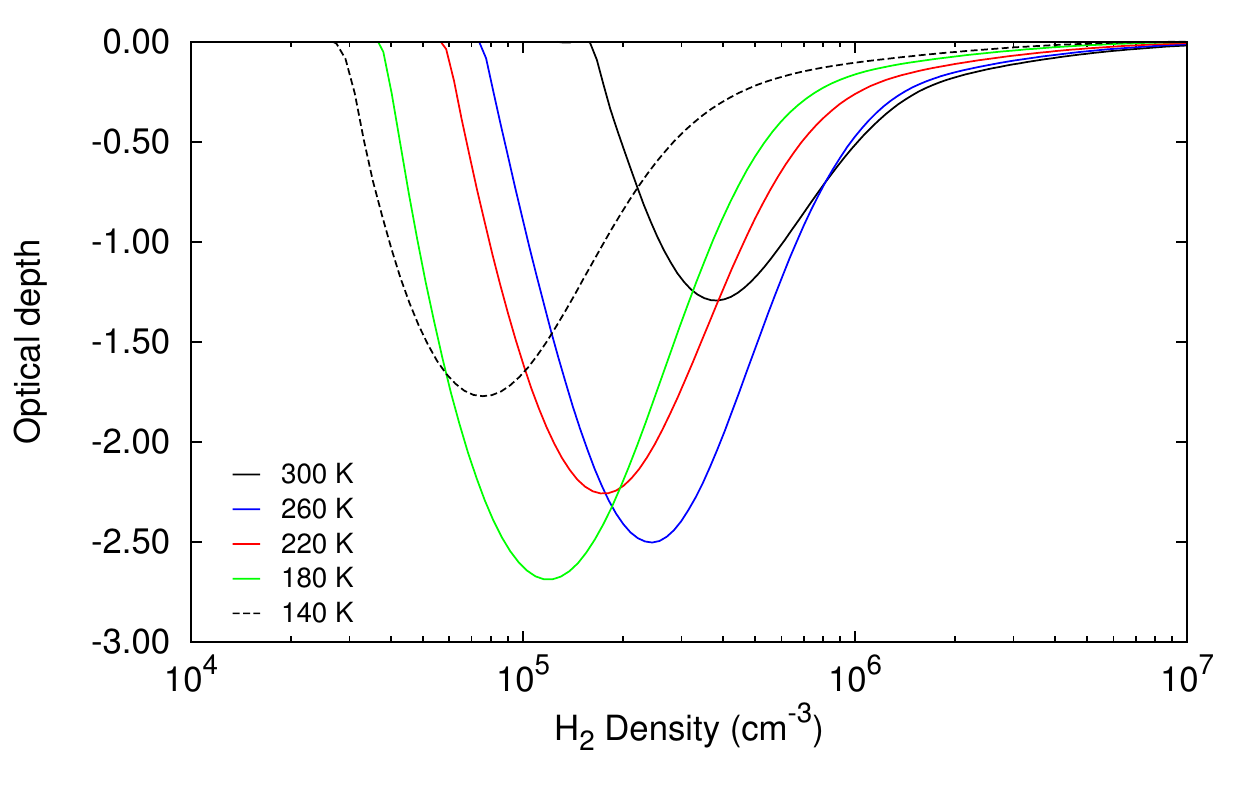}
     \caption{Comparison of the dependence of \taumax{} on \nhtwo{} for five values of \tk{}.}
     \label{fig:collonly}
   \end{figure}

In Figure \ref{fig:collonly} we therefore show the dependence of \taumax{} on \nhtwo{} for
$10^4~\mathrm{cm^{-3}} \leq n_{H_2} \leq 3 \times 10^7~\mathrm{cm^{-3}}$ for the five
values of \tk{} used in Fig. \ref{fig:tauvsncol1}. For each temperature the population
inversion occurs only above a ``threshold'' value of \nhtwo{} which increases with
increasing \tk{}. It is also seen that the density where \taumax{} is a maximum ranges
from about $7 \times 10^4~\mathrm{cm^{-3}}$ at \tk = 140 K to about $4\times
10^5~\mathrm{cm^{3}}$ at 300 K and that \taumax{} is a maximum for $T_k \sim$ 180 K. The
respective \htwo{} densities used in Fig. \ref{fig:tauvsncol1} are those where \taumax{}
has a maximum for the five values of \tk{}. It is also interesting to note that the
\htwo{} densities where \taumax{} is a maximum are significantly smaller for what is
found, for example, for OH masers \citep{pavlakis1996}.

\subsection{Radiative excitation: Dust infra-red radiation}

We examined the effect of the dust radiation field for different dust temperatures, and
values of $p$ and $W_d$ (equation \ref{eq:dust}) when collisions are absent. It can be
stated that no instance could be found where the dust emission on its own results in a
population inversion, not for the \gstate{} transition nor between any other levels.

   \begin{figure}[h]
     \centering
     \includegraphics[width=9cm]{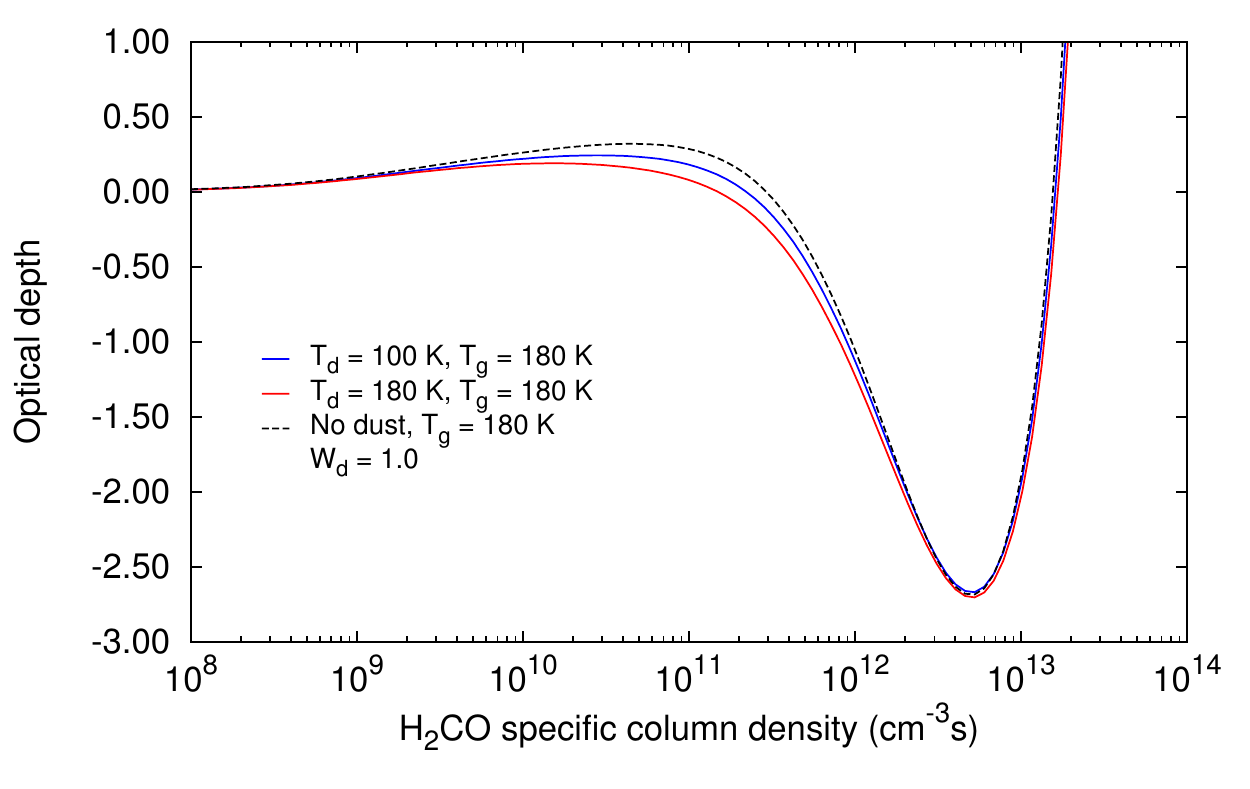}
     \caption{Comparison of the dependence of \taug{} on the \formaldehyde{} specific
       column density when excitation through collisions and the dust radiation field are
       included with the case when only collisional excitation is considered. All results
       are for \nhtwo = $1.3 \times 10^5~\mathrm{cm^{-3}}$.}
     \label{fig:collplusir01}
   \end{figure}

\begin{figure}[h]
  \centering
  \includegraphics[width=9cm]{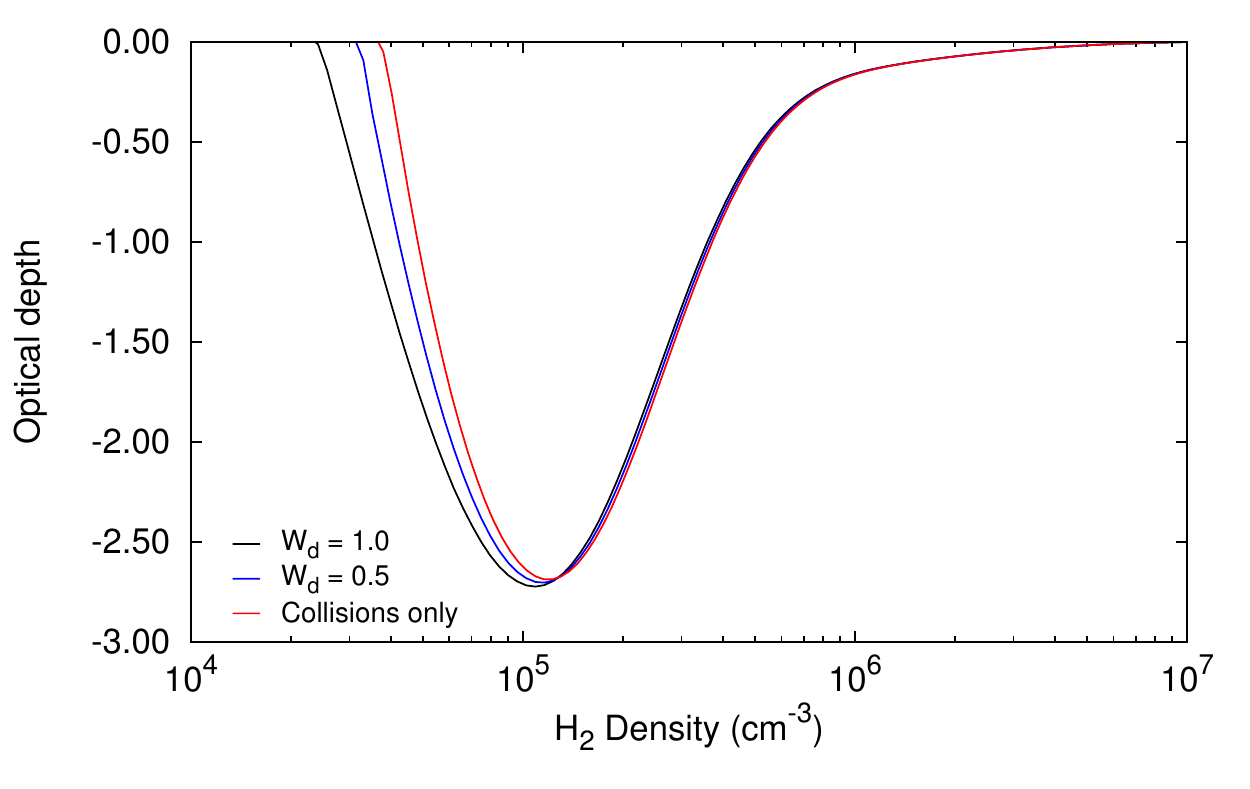}
  \caption{Comparison of the dependence of \taumax{} on \nhtwo{} when excitation through
    collisions and the dust radiation field is included with the case of pure collisional
    excitation is considered. In all three cases \tk = 180 K and \td = 100 K where the
    dust emission is included.}
  \label{fig:collplusir02}
\end{figure}

Although the dust radiation field in itself does not lead to a population inversion, it
nevertheless affects the level populations and therefore the population inversion of the
\gstate{} transition. The effect of the dust radiation field is small as is shown in
Fig. \ref{fig:collplusir01} where we compare \taug{} for the case of \tk{} = 180 K for
\td{} equal to 100 K and 180 K with pure collisional excitation when \tk{} = 180 K. In
Fig. \ref{fig:collplusir02} we show the dependence \taumax{} on \nhtwo{} for \tk{} = 180
K, \td{} = 100 K and for \wdust{} equal 0.5 and 1.0. The dust infrared radiation field is
seen to mostly affect \taumax{} for \nhtwo{} less than $10^5~\mathrm{cm^{-3}}$.

\subsection{Radiative excitation: Free-free continuum emission}

\begin{figure}[h]
  \centering
  \includegraphics[width=9cm]{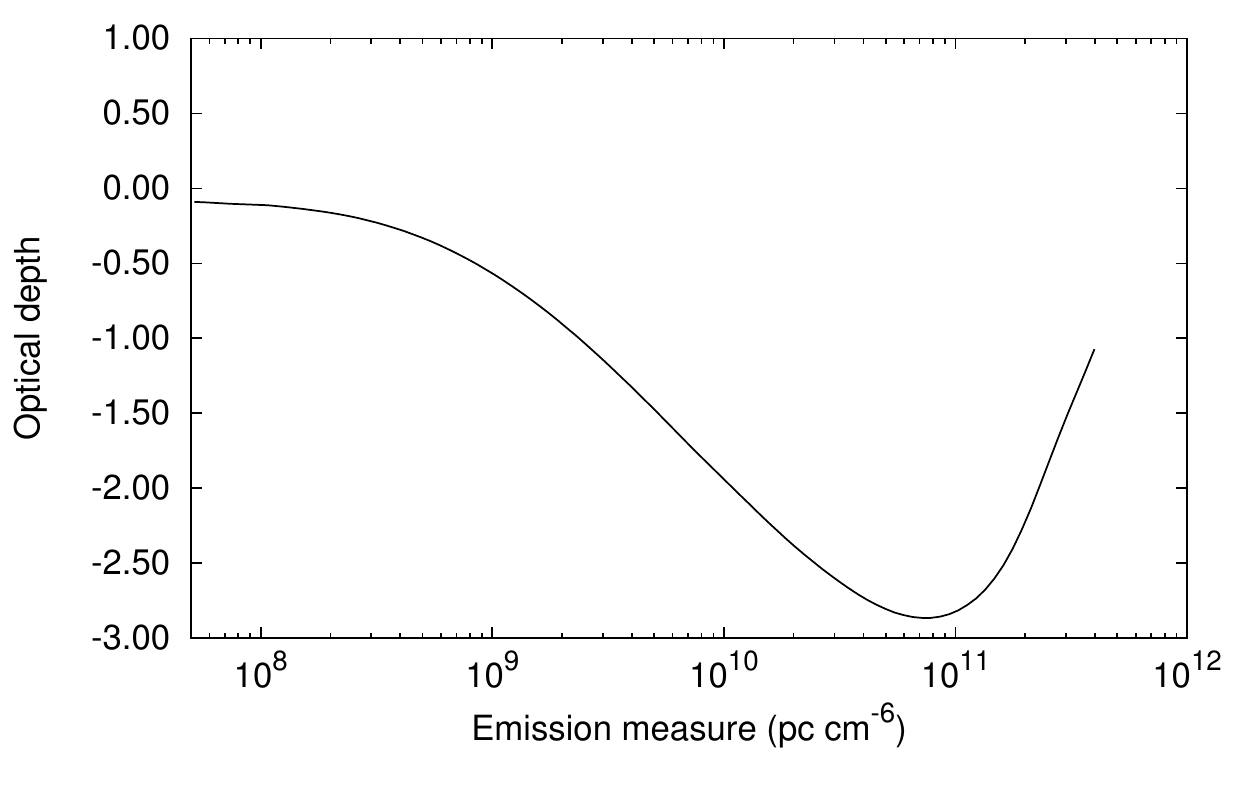}
  \caption{Dependence of \taumax{} on the emission measure for the case when collisions
    and the dust radiation field are ignored. The \htwo{} density used in this case was
    $10^5~\mathrm{cm^{-3}}$}
  \label{fig:onlyhii}
\end{figure}

In Fig. \ref{fig:onlyhii} we show the dependence of \taumax{} on the emission measure for
an \ion{H}{II} region with an electron temperature of $10^4$ K and a dilution factor,
\whii{} = 1 and where collisions were ignored. The range in emission measures from $5 \times
10^7$ to $4 \times 10^{11}$ $\mathrm{pc\,cm^{-6}}$ corresponds to turnover frequencies in
the radio continuum spectral energy distribution from about 4 GHz to 274 GHz. It is seen
that the free-free emission from the \ion{H}{II} region can cause a population inversion
for the \gstate{} transition. For emission measures greater than about $4 \times
10^{11}~\mathrm{pc\,cm^{-6}}$ \taug{} increases very rapidly to a value of 500 and remains
constant at that value. The maximum negative optical depth is found at an emission measure
of about $8 \times 10^{10}~\mathrm{pc\,cm^{-6}}$ which translates to a turnover frequency
of about 80 GHz.

\begin{figure}[h]
  \centering
  \includegraphics[width=9cm]{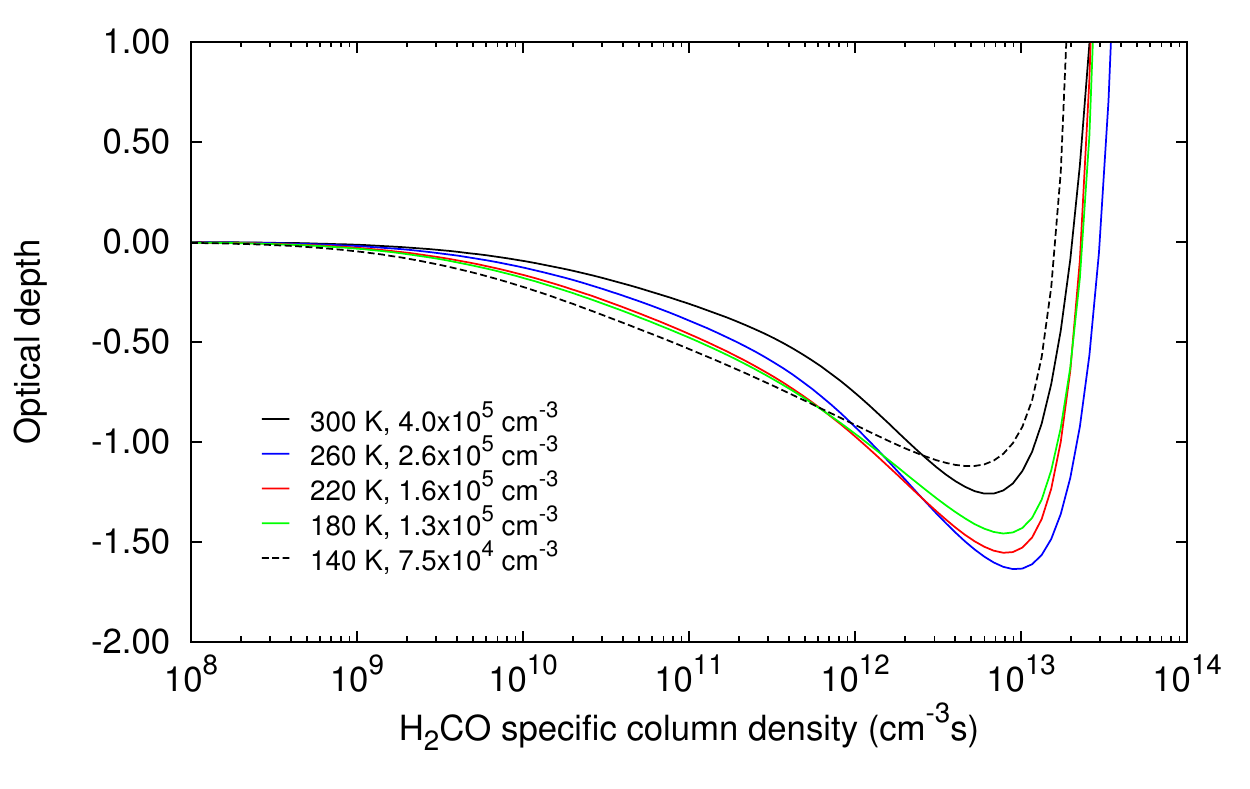}
  \caption{Graph showing the dependence of \taug{} on the \formaldehyde{} specific column
    density for five values of \tk{} in the presence of a free-free continuum radiation
    field. The emission measure used was $10^9~\mathrm{pc\,cm^{-6}}$ with \whii{} = 1. }
  \label{fig:hiipluscoll01}
\end{figure}

The effect of including collisional excitation is shown in
Fig. \ref{fig:hiipluscoll01}. The same combinations of \tk{} and \nhtwo{} as used in
Fig. \ref{fig:collonly} have been used.  The underlying dependence of \taug{} on the
specific column density as seen in Fig. \ref{fig:collonly} can also be seen in this case.
However, comparison with Fig. \ref{fig:collonly} shows that the free-free emission has a
significant effect in that an inversion of the \gstate{} transition now occurs for all
five combinations of \tk{} and \nhtwo{} already at a specific column density of
$10^8~\mathrm{cm^{-3}\,s}$ whereas this is not the case when collisions are the only
excitation mechanism. In fact, the effect on \taug{} is seen to be greatest for the lower
temperatures. The maximum negative values which \taug{} can have are, however, less than
in the case of pure collisional excitation. It should also be noted that the specific
column densities where \taug{} has a maximum negative value has shifted to larger values
in the presence of the free-free continuum radiation field than without it.

\begin{figure}[h]
  \centering 
  \includegraphics[width=9cm]{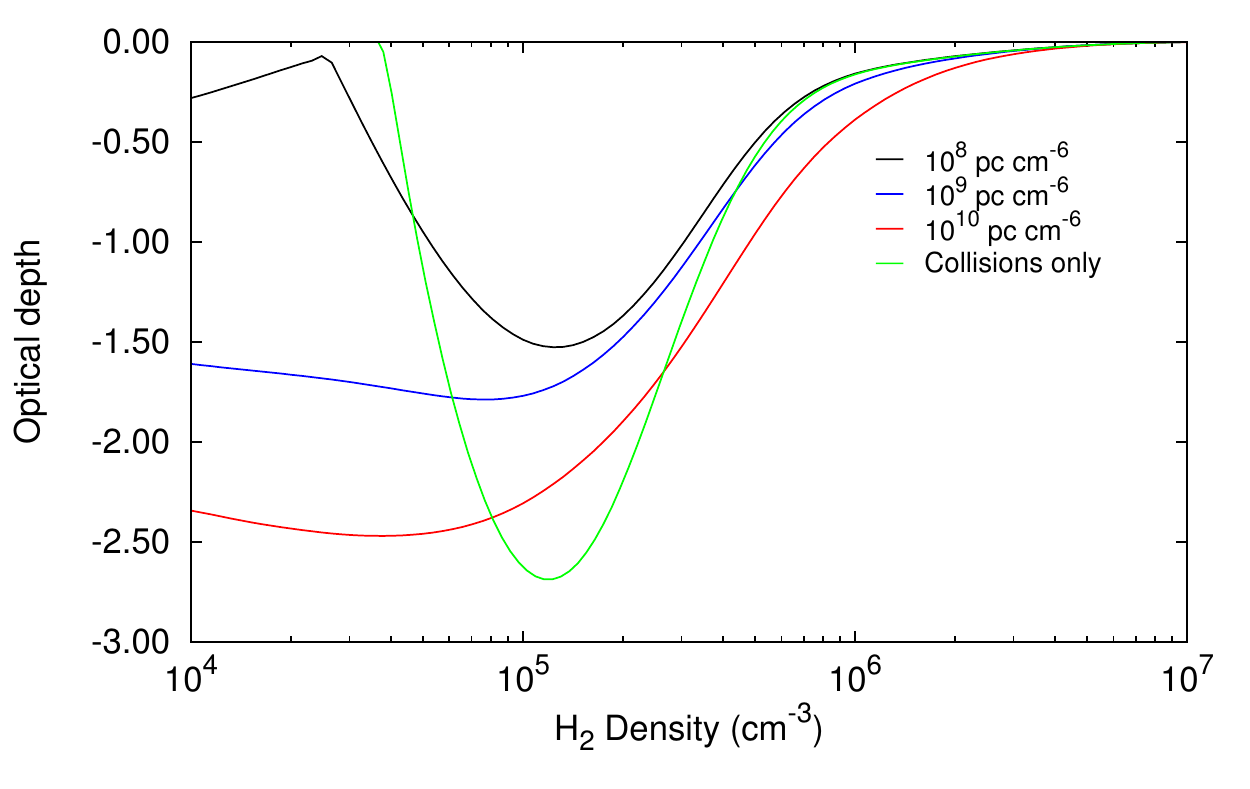}
  \caption{Comparison of \taumax{} as a function of \nhtwo{} for different emission
    measures. The gas kinetic temperature was taken as 180 K.}
  \label{fig:hiipluscoll02}
\end{figure}

With regard to the influence of the free-free continuum radiation field we lastly compare
in Fig. \ref{fig:hiipluscoll02} the dependence of \taumax{} on \nhtwo{} for emission
measures of $10^8~\mathrm{pc\,cm^{-6}}$, $10^9~\mathrm{pc\,cm^{-6}}$, and
$10^{10}~\mathrm{pc\,cm^{-6}}$. Also shown is the case for collisional excitation. In all
four cases we used \tk{} = 180 K. The free-free continuum radiation field is seen to
significantly affect the optical depth over most of the range of \nhtwo{} with the effect
being greatest toward lower densities where the inversion is exclusively due to that of
the free-free continuum radiation.

\subsection{Inversion of other transitions}

\begin{figure}[h]
  \centering 
  \includegraphics[width=9cm]{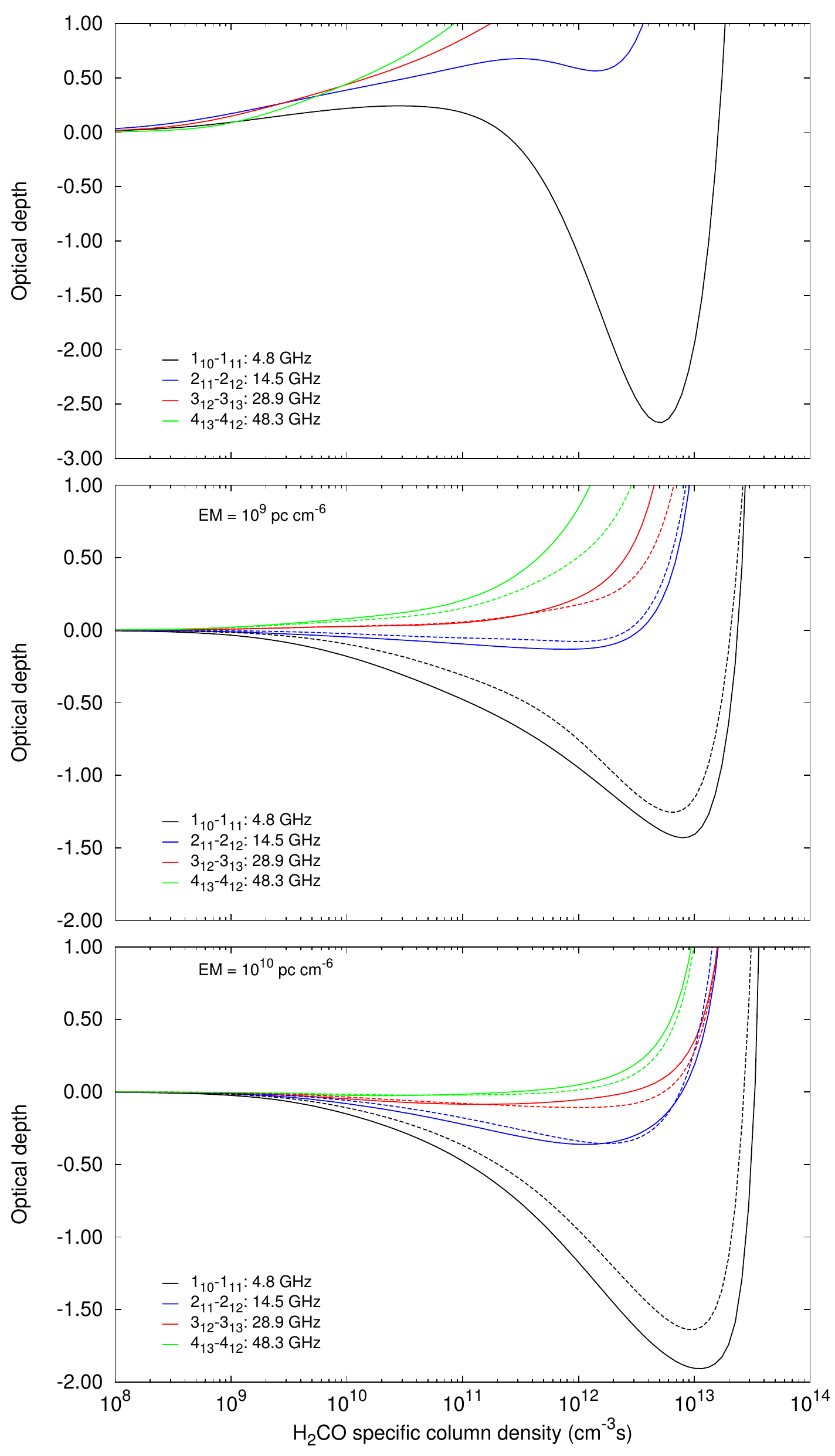}
  \caption{Comparison of optical depths for a number of transitions. The calculations were
    done for \td{} = 100 K, $W_d = 1.0$, and \whii{} = 1.0. The solid lines are for \tk{} = 180
    K with \nhtwo{} = $1.3 \times 10^5~\mathrm{cm^{-3}}$ and the dashed lines for \tk{} = 300
    K with \nhtwo{} = $4 \times 10^5~\mathrm{cm^{-3}}$. The emission measure is indicated in
  each of the panels.}
  \label{fig:manytrans}
\end{figure}

In Fig. \ref{fig:manytrans} we compare the dependence of the optical depth on the
\formaldehyde{} specific colum density for a number of transitions in order to investigate
the possibility of a population inversion for other transitions. We are in particular
interested in the \fstate{} transition since \citet{boland1981} predicted a detectable
maser line for this transition. In the upper panel we show the results for the case of
collisional excitation for the \gstate{} transition as well as for the \fstate{} (14.5
GHz), \sstate{} (28.9 GHz) and \tstate{} (48.3 GHz) transitions. Our choice for the last
three transitions is based on the slower spontaneous decay rates from the upper to the
lower levels for these transitions compared to other transitions. The gas kinetic
temperature was taken as 180 K and \nhtwo{} = $1.3 \times 10^5~\mathrm{cm^{-3}}$. In this
case only the \gstate{} transition shows a population inversion.

The middle and bottom panels show the results for more general cases where collisions and
the dust- and free-free continuum radiation fields were included. The middle panel is for
the case when the associated \ion{H}{II} region has an emission measure of
$10^9~\mathrm{pc\, cm^{-6}}$ and the bottom panel for an emission measure of
$10^{10}~\mathrm{pc\, cm^{-6}}$.  The \fstate{} transition is seen to show only a very
weak inversion while there are no inversions for the \sstate{} and \tstate{} transitions
for an emission measure of $10^9~\mathrm{pc\, cm^{-6}}$. For an emission measure of
$10^{10}~\mathrm{pc\,cm^{-6}}$ the \fstate{} transition shows a larger negative optical
depth while the \sstate{} transition shows only a very weak inversion. These results
suggests that apart from the \gstate{} transition, the \fstate{} and \sstate{} transitions
are inverted only in the presence of the free-free continuum radiation field and then only
for fairly large values of the emission measure. The optical depth for the 28.9 GHz maser
is, however, so small that it is unlikely that it will be detected while the 14.5 GHz
maser is expected to be detectable but significantly fainter than the 4.8 GHz maser.

\section{Discussion}

The results presented above show that the 4.8 GHz \formaldehyde{} masers can be pumped via
collisions with \htwo{} as well as radiatively by the free-free continuum emission from a
nearby ultra- or hypercompact \ion{H}{II} region but {\emph not} by the dust infrared
radiation field. Before discussing the implication of this result for the interpretation
of the \methanol{} and \formaldehyde{} flares in G37.55+0.20E is it necessary to consider
a few aspects of the results presented above.

Inspection of Figs. \ref{fig:tauvsncol1},~\ref{fig:collonly}, and \ref{fig:onlyhii} shows
that the maximum value for the optical depth for the \gstate{} transition lies between -2.5
and -3.0. This means that our calculations predict an amplification factor
($e^{-\tau_{4.8}}$) of at most 20. Even if the brightness temperature of the background
source is $10^4$ K it will therefore not be possible to explain maser brightness temperatures
greater than $2\times 10^5$ K. On the other hand, the model by \citet{boland1981} could
produce optical depths of -5 with which it is possible to explain brightness temperatures
of the order of $10^6$ K. The reason for this difference between the present calculations
and that of \citet{boland1981} is not clear. Within the framework of the present
calculations is it therefore possible to explain the fainter but not the brightest
Galactic 4.8 GHz \formaldehyde{} masers. 

A second point to be noted is that in the case of purely collisional excitation the
\formaldehyde{} specific column density where \taug{} is a maximum lies between about $5
\times 10^{12}~\mathrm{cm^{-3}\,s}$ and $10^{13}~\mathrm{cm^{-3}\,s}$. Although the aim of
the present calculations is not to discuss any particular source, the values of the
\formaldehyde{} specific column density where \taug{} reaches a maximum should be taken
note of as it might imply very large maser path lengths. For example, for \tk{} = 180 K
and \nhtwo{} = $1.3 \times 10^5~\mathrm{cm^{-3}}$ (see Fig. \ref{fig:tauvsncol1}), \taug{}
is a maximum at an \formaldehyde{} specific column density of $5 \times
10^{12}~\mathrm{cm^{-3}\,s}$. With the specific column density given by
$X_{H_2CO}n_{H_2}\ell/\Delta v$, the maser path length, $\ell$, is found to be 7.5 pc with
$\Delta v = 3~\mathrm{km\,s^{-1}}$ and $ X_{H_2CO} = 5 \times 10^{-7}$.  Such long path
lengths are certainly not realistic for masers associated with Galactic star forming
regions. It is interesting to note that \citet{araya2007} refer to having found a similar
result, ie. that very long, parsec scale, path lengths are required to explain the
observed brightness temperatures of the 4.8 GHz \formaldehyde{} masers.

To try to address this problem we note the following. First, in a recent study
\citet{mccauley2011} used the \sstate{} and \tstate{} transitions of o-\formaldehyde{} to
estimate densities in 23 well studied molecular cores. The \formaldehyde{} specific column
densities found by these authors lie between $1.4\times 10^8~\mathrm{cm^{-3}\,s}$ and
$5.1\times 10^{10}~\mathrm{cm^{-3}\,s}$ with an average of $2.9\times
10^9~\mathrm{cm^{-3}\,s}$. If the pumping is purely collisional then, refering to
Fig. \ref{fig:tauvsncol1}, it is seen that these specific column densities are well below
of what is required for a reasonable amplification factor for the \gstate{} transition. In
the presence of a very compact \ion{H}{II} region some inversion of the \gstate{} and
\fstate{} transitions would have been possible (bottom panel of Fig. \ref{fig:manytrans})
at these specific column densities. But even in this case the masers would have been very
weak. It would therefore appear as if masing of at least the \gstate{} transition requires
larger specific column densities than what seems to be typically found for
\formaldehyde{}. The need for larger specific column than what is normally found for
\formaldehyde{} in star forming regions {\emph might} be a reason for the rarity of the
\gstate{} \formaldehyde{} masers.

As already noted, the relatively large \formaldehyde{} specific column densities where the
amplification is significant implies unrealistically long maser path lengths. For a given
\htwo{} density, $\Delta v$, and specific column density, the only way to significantly
reduce the maser path length is to increase the relative abundance of \formaldehyde{}. In
the above example we assumed a relative abundance of $5 \times 10^{-7}$. However,
\citet{maret2004} found relative abundances for \formaldehyde{} of up to $6 \times
10^{-6}$ in the inner warm regions of low mass protostellar envelopes. Applying this
abundance to the above example would reduce the maser path length to 0.63 pc, which still
might be too large.  \citet{cragg2005} found that the 6.7 and 12.2 GHz \methanol{} masers
have their maximum brightness temperatures for a \methanol{} specific column density of
about $10^{13}~\mathrm{cm^{-3}\,s}$ which is comparable to what is required for the 4.8
GHz \formaldehyde{} maser. However, this result was found for \nhtwo{} =
$10^7~\mathrm{cm^{-3}}$ which is almost two orders of magnitude larger than the \htwo{}
densities where \taumax{} for the 4.8 GHz \formaldehyde{} masers is a maximum.  Using an
average relative abundance for \methanol{} of $10^{-6}$ \citep{maciel2013} and $\Delta v =
3 \times 10^5~\mathrm{cm\,s^{-1}}$, a \methanol{} maser path length of $3 \times
10^{17}~\mathrm{cm \sim 0.1~pc}$ is found.  If the path length of the 4.8 GHz
\formaldehyde{} maser is assumed to have this value, the \formaldehyde{} relative
abundance is required be approximately $4 \times 10^{-5}$.

Such a large relative abundance for \formaldehyde{} might seem unreasonable and it is
beyond the scope of the present work to make any statements in this regard. However, given
the rarity of \formaldehyde{} masers it might not be unreasonable to postulate rare
physical and chemical conditions in localized regions that can give rise to higher
relative abundances.  To explain the rarity of \formaldehyde{} masers \citet{hoffman2003}
suggested that the 4.8 GHz maser inversion might be due to a rare collisional
excitation. Our calculations have shown that inversion of the \gstate{} transition can be
achieved through collisional excitation with \htwo{} and that it is not necessary to
invoke a rare collisional mechanism. However, it seems as if it is still necessary to
invoke some rare conditions to explain the 4.8 GHz \formaldehyde{}
masers. \citet{gray2012} proposes that the rarity of the 4.8 GHz \formaldehyde{} masers
might be due to a short lifetime of these masers. A short lifetime alone, however, might
not be sufficient to explain the rarity of the masers. In addition to a short lifetime,
larger than average \formaldehyde{} abundances also seems to be required for the masers to
operate.

We now turn to the implications of the above results for the masers in G37.55+0.20. The
results presented above that the \gstate{} transition can be inverted collisionally via
\htwo{} and radiatively via the free-free continuum radiation but not through the dust
radiation field is contrary to the conclusion of \citet{araya2010} that the correlated
variability of the \methanol{} and \formaldehyde{} masers in G37.55+0.20 implies the same
excitation (pumping) mechanism for the two molecules.  If the pumping mechanisms of
\methanol{} and \formaldehyde{} are in fact different then any proposed mechanism should
be such that it can affect not only the dust temperature but also the gas kinetic
temperature and/or the free-free emission from an associated \ion{H}{II} region. As
already pointed out the \methanol{} and \formaldehyde{} maser flare profiles presented by
\citet{araya2010} are very similar.  Thus, a further constraint is that whatever mechanism
is proposed to simultaneously affect the pumping of the \methanol{} and \formaldehyde{}
masers, it should be such that the dust and gas temperatures and/or the properties of the
\ion{H}{II} region vary correctly in order to produce similar flare profiles. It is very
difficult to envisage a single mechanism or even a number of coupled processes that can be
responsible for such fine tuning, especially if the two masers are 2000 AU apart.

If the near simultaneous flaring of the \formaldehyde{} and \methanol{} masers cannot be
due to pumping effects, the only other possibility left is that it is due to changes in
the background seed photon flux. It has been postulated that the periodic \methanol{}
maser flares in G9.62+0.20E and G22.357+0.066 are due to changes in the free-free emission
from parts of the background \ion{H}{II} region against which they are projected
\citep{vanderwalt2011, szymczak2011}. Given that the flare profiles for the two masers in
G37.55+0.20 are very similar and also resemble that of the periodic masers in G9.62+0.20E
and G22.357+0.066, it is possible that the same underlying mechanism might be responsible
for the maser flaring in these three sources.

\section{Conclusions}
We presented the results of a numerical calculation to determine the physical conditions
under which a population inversion for the $1_{10} - 1_{11}$ transition of \formaldehyde{}
can occur.  The underlying aim of these calculations was to interpret the near
simultaneous flaring of the 6.7 GHz \methanol{} and 4.8 GHz \formaldehyde{} masers in
G37.55+0.20. Considering the results presented above we conclude the following:
\begin{itemize}
\item The \gstate{} transition can be inverted via collisional excitation with \htwo{} as
  well as by the free-free continuum of \ion{H}{II} regions with emission measures greater
  than about $5 \times 10^7~\mathrm{pc\, cm^{-6}}$.
\item The dust infrared emission does affect the level populations but does not result in
  a population inversion of the \gstate{} transition nor of any other possible masing
  transitions.
\item The maximum amplification found within the framework of our calculations
  is only about 20 which cannot explain the brightest 4.8 GHz \formaldehyde{}
  masers. 
\item If collisions are the only excitation mechanism a noteable amplification for the 4.8
  GHz masers exists only for \formaldehyde{} specific column densities above about $5
  \times 10^{11}~\mathrm{cm^{-3}\,s}$. These specific column densities are significantly
  larger than what is derived from thermal line emission of \formaldehyde{}.
\item To avoid unrealistically large maser path lengths it is necessary to postulate
  higher than average relative abundances for \formaldehyde{} in the regions where the 4.8
  GHz masers operate.
\item The \fstate{} and \sstate{} transitions are inverted only in the presence
  of the free-free continuum radiation field of very compact \ion{H}{II}
  regions.  
\item Due to the different pumping mechanisms of the \formaldehyde{} and \methanol{}
  masers it is unlikely that the near simultaneous flaring of the \methanol{} and
  \formaldehyde{} masers in G37.55+0.20 is due to changes in the pumping of the
  masers. Instead, as an alternative it is suggested that the flaring is due to a single
  process that affects the flux of the background seed photon flux for both of the masers
  even though they are 2000 AU apart.
\end{itemize} 
\begin{acknowledgements}

I would like to thank Melvin Hoare and Sharmila Goedhart for valuable comments on an
earlier version of the manuscript as well as an anonymous referee and Malcolm Walmsley for
useful comments. This work was supported by the National Research Foundation under grant
number 2053475.
\end{acknowledgements}
\bibliographystyle{aa} 
\bibliography{ref.bib}

\begin{thebibliography}{26}
\expandafter\ifx\csname natexlab\endcsname\relax\def\natexlab#1{#1}\fi

\bibitem[{{Araya} {et~al.}(2007){Araya}, {Hofner}, {Kurtz}, {Linz}, {Sewilo},
  {Olmi}, \& {Watson}}]{araya2007}
{Araya}, E., {Hofner}, P., {Kurtz}, S., {et~al.} 2007, in IAU Symposium, Vol.
  242, IAU Symposium, ed. {J.~M.~Chapman \& W.~A.~Baan}, 140--141

\bibitem[{{Araya} {et~al.}(2010){Araya}, {Hofner}, {Goss}, {Kurtz}, {Richards},
  {Linz}, {Olmi}, \& {Sewi{\l}o}}]{araya2010}
{Araya}, E.~D., {Hofner}, P., {Goss}, W.~M., {et~al.} 2010, \apjl, 717, L133

\bibitem[{{Baan} {et~al.}(1986){Baan}, {Guesten}, \& {Haschick}}]{baan1986}
{Baan}, W.~A., {Guesten}, R., \& {Haschick}, A.~D. 1986, \apj, 305, 830

\bibitem[{{Boland} \& {de Jong}(1981)}]{boland1981}
{Boland}, W. \& {de Jong}, T. 1981, \aap, 98, 149

\bibitem[{{Cragg} {et~al.}(2002){Cragg}, {Sobolev}, \& {Godfrey}}]{cragg2002}
{Cragg}, D.~M., {Sobolev}, A.~M., \& {Godfrey}, P.~D. 2002, \mnras, 331, 521

\bibitem[{{Cragg} {et~al.}(2005){Cragg}, {Sobolev}, \& {Godfrey}}]{cragg2005}
{Cragg}, D.~M., {Sobolev}, A.~M., \& {Godfrey}, P.~D. 2005, \mnras, 360, 533

\bibitem[{{Goedhart} {et~al.}(2003){Goedhart}, {Gaylard}, \& {van der
  Walt}}]{goedhart2003}
{Goedhart}, S., {Gaylard}, M.~J., \& {van der Walt}, D.~J. 2003, \mnras, 339,
  L33

\bibitem[{{Goedhart} {et~al.}(2007){Goedhart}, {Gaylard}, \& {van der
  Walt}}]{goedhart2007}
{Goedhart}, S., {Gaylard}, M.~J., \& {van der Walt}, D.~J. 2007, in IAU
  Symposium, Vol. 242, IAU Symposium, ed. {J.~M.~Chapman \& W.~A.~Baan},
  97--101

\bibitem[{{Goedhart} {et~al.}(2009){Goedhart}, {Langa}, {Gaylard}, \& {van der
  Walt}}]{goedhart2009}
{Goedhart}, S., {Langa}, M.~C., {Gaylard}, M.~J., \& {van der Walt}, D.~J.
  2009, \mnras, 398, 995

\bibitem[{{Gray}(2012)}]{gray2012}
{Gray}, M. 2012, {Maser Sources in Astrophysics} (Cambridge University Press)

\bibitem[{{Hoffman} {et~al.}(2003){Hoffman}, {Goss}, {Palmer}, \&
  {Richards}}]{hoffman2003}
{Hoffman}, I.~M., {Goss}, W.~M., {Palmer}, P., \& {Richards}, A.~M.~S. 2003,
  \apj, 598, 1061

\bibitem[{{Kreyszig}(1979)}]{kreyszig}
{Kreyszig}, E. 1979, {Advanced Engineering Mathematics} (Wiley)

\bibitem[{{Maciel}(2013)}]{maciel2013}
{Maciel}, W.~J. 2013, {Astrophysics of the Interstellar Medium} (Springer)

\bibitem[{{Maret} {et~al.}(2004){Maret}, {Ceccarelli}, {Caux}, {Tielens},
  {J{\o}rgensen}, {van Dishoeck}, {Bacmann}, {Castets}, {Lefloch}, {Loinard},
  {Parise}, \& {Sch{\"o}ier}}]{maret2004}
{Maret}, S., {Ceccarelli}, C., {Caux}, E., {et~al.} 2004, \aap, 416, 577

\bibitem[{{McCauley} {et~al.}(2011){McCauley}, {Mangum}, \&
  {Wootten}}]{mccauley2011}
{McCauley}, P.~I., {Mangum}, J.~G., \& {Wootten}, A. 2011, \apj, 742, 58

\bibitem[{{Mehringer} {et~al.}(1994){Mehringer}, {Goss}, \&
  {Palmer}}]{mehringer1994}
{Mehringer}, D.~M., {Goss}, W.~M., \& {Palmer}, P. 1994, \apj, 434, 237

\bibitem[{{Menten}(2012)}]{menten2012}
{Menten}, K.~M. 2012, in IAU Symposium, Vol. 287, IAU Symposium, ed. R.~S.
  {Booth}, W.~H.~T. {Vlemmings}, \& E.~M.~L. {Humphreys}, 506--515

\bibitem[{{Pavlakis} \& {Kylafis}(1996)}]{pavlakis1996}
{Pavlakis}, K.~G. \& {Kylafis}, N.~D. 1996, \apj, 467, 309

\bibitem[{{Sch{\"o}ier} {et~al.}(2005){Sch{\"o}ier}, {van der Tak}, {van
  Dishoeck}, \& {Black}}]{lamda2005}
{Sch{\"o}ier}, F.~L., {van der Tak}, F.~F.~S., {van Dishoeck}, E.~F., \&
  {Black}, J.~H. 2005, \aap, 432, 369

\bibitem[{{Sobolev} {et~al.}(1997){Sobolev}, {Cragg}, \&
  {Godfrey}}]{sobolev1997}
{Sobolev}, A.~M., {Cragg}, D.~M., \& {Godfrey}, P.~D. 1997, \aap, 324, 211

\bibitem[{{Sobolev} \& {Deguchi}(1994)}]{sobolev1994}
{Sobolev}, A.~M. \& {Deguchi}, S. 1994, \aap, 291, 569

\bibitem[{{Szymczak} {et~al.}(2011){Szymczak}, {Wolak}, {Bartkiewicz}, \& {van
  Langevelde}}]{szymczak2011}
{Szymczak}, M., {Wolak}, P., {Bartkiewicz}, A., \& {van Langevelde}, H.~J.
  2011, \aap, 531, L3

\bibitem[{{Thaddeus}(1972)}]{thaddeus1972}
{Thaddeus}, P. 1972, \apj, 173, 317

\bibitem[{{van der Walt}(2011)}]{vanderwalt2011}
{van der Walt}, D.~J. 2011, \aj, 141, 152

\bibitem[{{van der Walt} {et~al.}(2009){van der Walt}, {Goedhart}, \&
  {Gaylard}}]{vanderwalt2009}
{van der Walt}, D.~J., {Goedhart}, S., \& {Gaylard}, M.~J. 2009, \mnras, 398,
  961

\bibitem[{{Wiesenfeld} \& {Faure}(2013)}]{wiesenfeld2013}
{Wiesenfeld}, L. \& {Faure}, A. 2013, \mnras, 432, 2573

\end{thebibliography}
\end{document}